\documentclass[aps,pra,twocolumn,nofootinbib,twocolumn,floatfix,showpacs,superscriptaddress]{revtex4-2}
\usepackage[utf8]{inputenc}
\usepackage{amsmath,graphicx,epsfig,amsthm,color,bm,braket}
\usepackage{pifont,bbding,amssymb,soul,mathrsfs}
\usepackage{indentfirst}
\usepackage{times,dsfont,units}
\usepackage{appendix, comment}

\usepackage{ulem}

\usepackage{dcolumn}

\usepackage[hyperindex,breaklinks,colorlinks=true,citecolor=blue,urlcolor=blue]{hyperref}

\begin{document}
\title{Experimental High-Accuracy and Broadband Quantum Frequency Sensing via Geodesic Control}
\author{Si-Qi Chen}
\email{These authors contributed equally to this work.}
\affiliation{School of Physics, State Key Laboratory of Crystal Materials, Shandong University, Jinan 250100, China}

\author{Qi-Tao Duan}
\email{These authors contributed equally to this work.}
\affiliation{School of Physics, State Key Laboratory of Crystal Materials, Shandong University, Jinan 250100, China}

\author{Teng Li}
\affiliation{School of Physics, State Key Laboratory of Crystal Materials, Shandong University, Jinan 250100, China}

\author{He Lu}
\email{luhe@sdu.edu.cn}
\affiliation{School of Physics, State Key Laboratory of Crystal Materials, Shandong University, Jinan 250100, China}

\begin{abstract}
Accurate frequency estimation of oscillating signals over a broad bandwidth is a central task in quantum sensing, yet it is often compromised by spurious responses to higher-order harmonics in realistic multi-frequency environments. Here we experimentally demonstrate a high-accuracy and broadband quantum frequency sensing protocol based on geodesic control, implemented using the electron spin of a single nitrogen–vacancy center in diamond. By engineering an intrinsically single-frequency response, geodesic control enables bias-free frequency estimation with strong suppression of harmonic-induced systematic errors across a wide spectral range spanning from the megahertz to the gigahertz regime. Furthermore, by incorporating synchronized readout, we achieve millihertz-level frequency resolution under noisy signal conditions. Our results provide systematic experimental benchmarking of geodesic control for quantum frequency sensing and establish it as a practical approach for high-accuracy metrology in realistic environments.
\end{abstract}
\maketitle

Quantum sensing exploits uniquely quantum resources, such as coherence and entanglement, to achieve measurement sensitivities beyond classical noise limits, enabling powerful capabilities for both fundamental science and emerging technologies~\cite{Degen2017RMP}. A pivotal task in this field is the accurate estimation of the frequency of oscillating (AC) signals, which underpins applications ranging from radar and wireless communication~\cite{Georges1996nature,Hui2019nature,Zhang2020NC,Patel2024PRApl} to nanoscale nuclear magnetic resonance~\cite{Staudacher2013science,DeVience2015NatureNano,Rugar2015NC,Loretz2015APL,Schmitt2017science,Kehayias2017NC,Glenn2018nature,Bucher2020PRX} and entanglement detection of microwave photons~\cite{Bozyigit2011NP,Gasparinetti2017PRL}. Despite significant progress, achieving robust and high-accuracy frequency estimation in realistic environments remains a major challenge, primarily due to multi-frequency noise and systematic errors inherent to the sensing protocols themselves.

Dynamical decoupling (DD) sequences constitute one of the most widely used approaches for AC signal sensing, as they impart frequency selectivity to the quantum sensor while suppressing broadband environmental noise~\cite{Kotler2011nature,Lange2011PRL,Alvarez2011PRL}. When combined with complementary techniques such as quantum mixers~\cite{Wang2022PRX}, heterodyne detection~\cite{Chu2021PRApplied,Meinel2021NC} and synchronized readout technique~\cite{Boss2017science,Schmitt2017science,Glenn2018nature}, DD-based sensing schemes can access wide frequency ranges and achieve high spectral resolution and precision. However, in practical sensing scenarios, target AC signals are typically embedded in complex, multi-frequency noise backgrounds. Standard DD sequences, including CPMG~\cite{Carr1954PR,Meiboom1958RSI} and XY families~\cite{Terry1990JMR}, while optimized for a target frequency $\omega_s$, intrinsically exhibit non-negligible sensitivity to higher-order odd harmonics at $\omega_n = k\omega_s$ ($k=3,5,\ldots$)~\cite{Loretz2015PRX,Degen2017RMP}. These spurious harmonic responses lead to systematic biases and and severely limit sensing accuracy, constituting a fundamental obstacle for quantum frequency metrology under realistic conditions.

Motivated by this limitation, recent theoretical work has proposed engineering the sensor’s control trajectory itself to eliminate unwanted harmonic responses. In particular, geodesic control has been shown to enable sensing protocols with an intrinsically single-frequency response, suppressing spurious harmonics through the geometry of quantum evolution rather than through additional filtering or post-processing~\cite{Wang2016PRA,Zeng2024PRL}. This theoretical framework points toward a fundamentally different route to robust frequency estimation, but its experimental feasibility, resilience to realistic noise, and compatibility with broadband and high-resolution sensing techniques have remained open questions.

In this work, we experimentally demonstrate a high-accuracy and broadban quantum frequency sensing protocol based on geodesic control, using the electron spin of a single nitrogen–vacancy (NV) center in diamond as a quantum sensor. As shown in Fig.~\ref{fig:1}~(a), geodesic control shapes the modulation spectrum of the quantum sensor into an effectively tunable single-frequency response, thereby strongly suppressing spurious harmonic resonances. We quantitatively characterize this frequency selectivity by reconstructing the modulation spectrum and monitoring the sensor-state evolution under controlled multi-frequency noise, and directly compare its performance with conventional DD-based schemes. Our experiments demonstrate accurate and bias-free frequency estimation is achievable for AC signals in the megahertz~(MHz) regime, even in the presence of higher-order harmonic noise. By combining geodesic control with heterodyne detection, we extend the accessible sensing bandwidth into the gigahertz~(GHz) regime without sacrificing accuracy. Furthermore, by incorporating synchronized readout, we achieve millihertz-level frequency resolution while maintaining strong suppression of noise-induced spurious responses. Compared with conventional XY and CPMG protocols, geodesic control exhibits a marked improvement in both accuracy and robustness, establishing a practical and scalable framework for high-resolution and broadband quantum frequency sensing in realistic environments. 

\begin{figure*}[ht!bp]
\includegraphics[width=\linewidth]{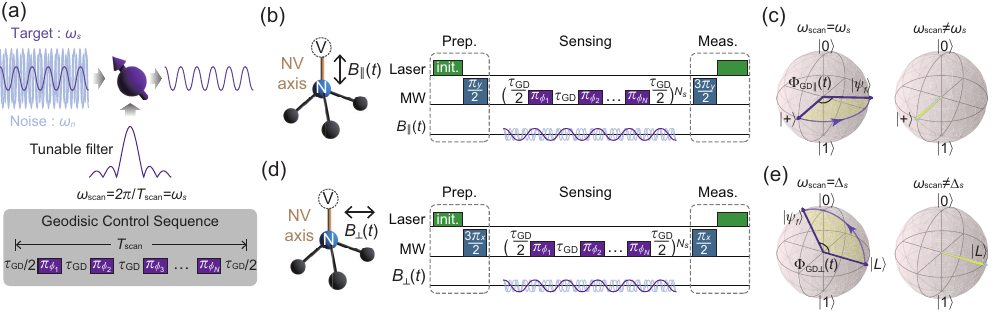}
\caption{(a) Schematic of frequency sensing using a single NV electron spin. The geodesic control of the electron spin acts as a tunable spectral filter, enabling selective detection of the target frequency $\omega_s$ in the presence of a noisy environment. (b) The pulse sequence exploiting the parallel component $B_\parallel(t)$ to sense the low-frequency signals in the MHz regime. (c) Accumulated phase $\Phi_{\text{GD}_\parallel}(t)$ as a function of the scan frequency, exhibiting strong spectral selectivity with a maximum response at $\omega_\text{scan}=\omega_s$. (d) Heterodyne pulse sequence exploiting the perpendicular component $B_\perp(t)$ to enable frequency sensing in the GHz regime. (e) Accumulated phase $\Phi_{\text{GD}_\perp}(t)$ as a function of the scan frequency, exhibiting strong spectral selectivity with a maximum response at $\omega_\text{scan}=\Delta_s$.
\label{fig:1}}
\end{figure*}

For the electron spin associated with a single NV center in diamond, the electronic ground state forms a spin triplet $^3$A with sublevels $\ket{m_s=0}$ and $\ket{m_s=\pm1}$. The NV-center Hamiltonian is $H_{NV}=DS_z^2+\gamma_e B_0 S_z$, where $D=2\pi\times$~2.87~GHz is the zero-field splitting and $\gamma_e=2\pi\times 28~\mathrm{GHz/T}$ is the electron gyromagnetic ratio. A static field $B_0\approx500$~Gs, applied along the NV axis, lifts the degeneracy between $\ket{m_s=\pm1}$, introducing a Zeeman splitting of $2\gamma_e B_0$. For sensing, we restrict the dynamics to the $\{\ket{m_s=0}, \ket{m_s=-1}\}$ subspace, defining the qubit states $\ket{0}$ and $\ket{1}$, respectively. In this basis, the Hamiltonian reduces to $H_0=-\omega_0 \sigma_z/2$ with $\omega_0 = D - \gamma_e B_0\approx2\pi\times1.47$~GHz. We consider an AC magnetic field $B(t)$ with components parallel and perpendicular to the NV axis
\begin{eqnarray}\label{Eq:ACparallel}
    B_\parallel(t)=b_s^\parallel\cos(\omega_st) + \sum_l b_{n,l}^\parallel \cos(\omega_{n,l} t),\\\label{Eq:ACperp}
    B_\perp(t)=b_s^\perp\cos(\omega_st) + \sum_l b_{n,l}^\perp \cos(\omega_{n,l} t),
\end{eqnarray}
where $b_s^\parallel$~($b_s^\perp$), $\omega_s$ denote the amplitude and angular frequency of the target signal projected parallel~(perpendicular) to the NV axis, respectively. $b_{n,l}^\parallel$~($b_{n,l}^\perp$) and $\omega_{n,l}$ characterize the amplitudes and frequencies of the noise components.

For $B(t)$ in the MHz regime, we only consider its parallel component $B_\parallel(t)$ acting on NV center. As shown in Fig.~\ref{fig:1}~(b), the electron spin is prepared on $\ket{\psi_0^\parallel}=\ket{+}=(\ket{0}+\ket{1})/\sqrt{2}$, followed by the application of the GD$_\parallel$ sequence, which consists of $N_s$ repetitions of geodesic pulse block
\begin{equation} \label{Geodesic sequence}
    \left(\frac{\tau_{\text{GD}}}{2}-\pi_{\phi_1}-\tau_{\text{GD}} -\pi_{\phi_2}-\cdots-\tau_{\text{GD}} -\pi_{\phi_N}-\frac{\tau_{\text{GD}}}{2}\right)^{N_s}.
\end{equation}
A geodesic pulse block consists of $N$ sequential $\pi$ rotations applied over a total scanning time $T_\text{scan}=N(\tau_\text{\text{GD}}+t_{\pi})$, where $\tau_{\mathrm{GD}}$ denotes the free-evolution interval between adjacent $\pi$ pulses. The $j$-th $\pi$ pulse, denoted by $\pi_{\phi_j}$, has a duration of $t_\pi$ and is centered at time $T_j = T_{\mathrm{scan}}(2j-1)/2N, j = 1, 2, \dots, N$. Each pulse implements a $\pi$ rotation about the axis $(\sin\phi_j,0,-\cos\phi_j)$ in $x$-$z$ plane, where the phase $\phi_j$ varies in time according to $\phi_j = 2\pi T_j/T_{\mathrm{scan}}$. The Hamiltonian of GD$_\parallel$ is thus given by
\begin{equation} \label{Eq:GDy_control}
    H_{\text{GD}_\parallel}=\frac{\Omega_{\text{GD}}(t)}{2}(-\cos\phi_{\text{GD}}(t)\sigma_z+\sin\phi_{\text{GD}}(t)\sigma_x).
\end{equation}
The Rabi amplitude of $\pi_{\phi_j}$ pulse is set at $\Omega_{\text{GD}}(t) = \pi / t_{\pi}$, and the phase parameter is fixed at $\phi_{\text{GD}}(t)=\phi_j=2\pi T_j/T_{\text{scan}}$. In the interaction picture defined by $H_{\text{GD}_\parallel}$~\cite{Choi2020PRX}, the total sensing Hamiltonian takes the form
\begin{equation} \label{Eq:effective}
    \widetilde{H}_\parallel=F_{\text{GD}_\parallel}(t)  B_\parallel(t)\frac{\sigma_{z}}{2},
\end{equation}
where $F_{\text{GD}_\parallel}(t)$ is the modulation function determined by the GD$_\parallel$ control sequence. In the limit of large $N$, the ideal modulation function approaches a single-frequency form, i.e., $F_{\text{GD}_\parallel}(t)\approx \cos(2\pi t /T_{\text{scan}})$, and we define the scanning frequency as $\omega_{\text{scan}}=2\pi/T_{\text{scan}}$.

\begin{figure*}[ht!bp]
\includegraphics[width=\linewidth]{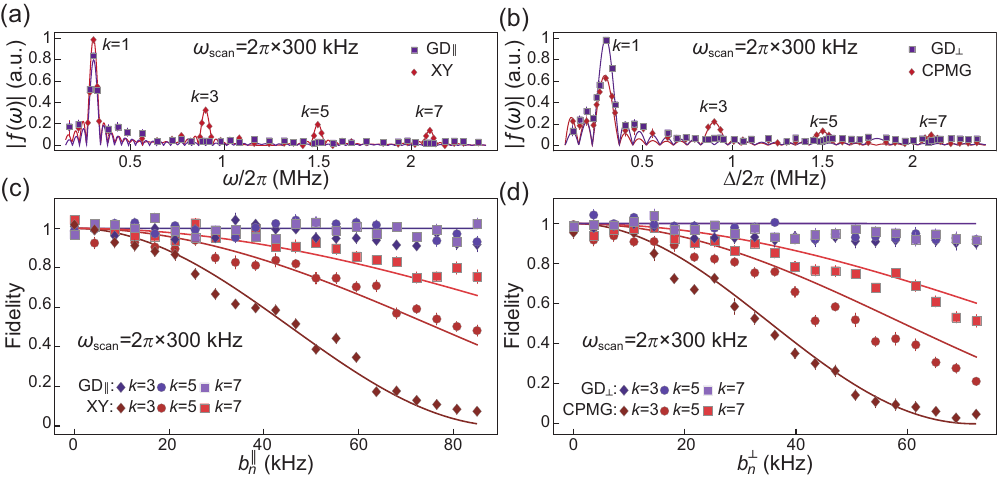}
\caption{The Fourier components of the modulation functions under finite-width control pulses for (a) MHz signal measurement and (b) GHz signal measurement. a.u., arbitrary units. The solid lines come from the numerical simulation. The diamonds, squares represent the experimental results of conventional and geodesic schemes, respectively. Robustness of the (c) MHz- and (d) GHz-regime sensing protocols against noise signals at the 3rd, 5th and 7th harmonics with varying amplitudes. The diamonds, dots and squares represent the experimental results obtained by adding a single noise component at the $k=3$, $k=5$ and $k=7$ harmonics, respectively. The solid lines represent the results from numerical simulation. The error bars represent the uncertainties estimated by adding Poisson noise to the experimental data.\label{fig:2}}
\end{figure*}

Governed by the effective Hamiltonian $\widetilde{H}_\parallel$ in Eq.~(\ref{Eq:effective}), the initial state $\ket{\psi_0^\parallel}$ undergoes a rotation about the $z$ axis by an angel $\Phi_{\text{GD}_\parallel}(t)$, which depends on the scan frequency $\omega_\text{scan}$ as
\begin{equation}
\begin{split}
    \Phi_{\text{GD}_\parallel}(t)&=\int_0^tF_{\text{GD}_\parallel}(t^\prime) B_\parallel(t^\prime)\,dt^\prime\\
    &\approx\int_0^t \Bigg\{\frac{b_s}{2}\cos\left[(\omega_s-\omega_{\text{scan}})t^\prime\right]\\
    &+ \sum_l \frac{b_{n,l}}{2}\cos\left[(\omega_{n,l}-\omega_{\text{scan}}) t^\prime\right]\Bigg\}\,dt^\prime.\\
\end{split}
\end{equation}
Notably, the accumulated phase $\Phi_{\text{GD}_\parallel}(t)$ exhibits strong frequency selectivity, i.e., it builds up constructively when the scan frequency matches the target frequency  $\omega_{\text{scan}}=\omega_s$, whereas it is strongly suppressed for $\omega_{\text{scan}}\neq\omega_s$, as illustrated in Fig.~\ref{fig:1}~(c). This frequency resonance can be read out by projecting the final state onto $\ket{+}$, yielding the measurement probability  $P_+=\frac{1}{2}\left[1+\cos(\Phi_{\text{GD}_\parallel}(t))\right]$. Consequently, when $\omega_{\text{scan}}$ is tuned to the target frequency $\omega_s$, contributions from noise components at other frequencies are efficiently filtered out, enabling high-accuracy determination of the target signal frequency.

Indeed, the scanning frequency $\omega_\text{scan}$ can, in principle,be extended to the GHz regime. However, this requires the $\pi$-pulse duration $t_{\pi}$ to be sufficiently short, which in practice is limited by the available control field power. This limitation can be circumvented by employing a heterodyne technique, in which we exploit the component $B_\perp(t)$ perpendicular to the NV axis as shown in Fig.~\ref{fig:1}~(d). 
By transforming into a rotating frame defined by the qubit splitting $\omega_0$, the fast-oscillating transverse field is converted into an effective low-frequency signal characterized by the detunings
$\Delta_s = \omega_s - \omega_0$ and $\Delta_{n,l} = \omega_{n,l} - \omega_0$~\cite{Meinel2021NC,Chu2021PRApplied}.
To sense $\Delta_s$, the electron state is first prepared on $\ket{\psi_0^\perp}=\ket{L}=(\ket{0}+i\ket{1})/\sqrt{2}$, and the same geodesic control sequence~(Eq.~(\ref{Geodesic sequence})) is applied, with each $\pi_{\phi_j}$ corresponding to a $\pi$ rotation about the axis $(\cos\phi_j, -\sin\phi_j, 0)$ in $x$-$y$ plane (denoted as GD$_\perp$). Under this control, the initial state $\ket{\psi_0^\perp}$ undergoes a rotation round $x$ axis by an angle of $\Phi_{\text{GD}_\perp}(t)$ as shown in Fig.~\ref{fig:1}~(e). Analogous to the parallel case, $\Phi_{\text{GD}_\perp}(t)$ exhibits strong frequency selectivity as well, i.e., it accumulates constructively when $\omega_{\text{scan}}=\Delta_s$, while remaining strongly suppressed for off-resonant detunings. This accumulated phase is read out by projecting the final state, yielding probability of $P_L=\frac{1}{2}[1+\cos(\Phi_{\text{GD}_\perp}(t))]$. More details of geodesic control are provided in the Supplementary Materials.

Experimentally, the electron spin is initialized, manipulated and readout using a home-build confocal microscope and microwave antenna~\cite{Chen2025PRApplied, Chen2025PRA, Duan2025Arxiv}. To characterize the noise-suppression property of GD sequences, we experimentally reconstructed the Fourier components $|f(\omega)|$ of their respective modulation functions. The scanning frequencies $\omega_\text{scan}$ for both GD$_\parallel$ and GD$_\perp$ are fixed at $2\pi\times0.3$~MHz. For the GD$_\parallel$ sequence, the signal frequency $\omega_s$ is swept from $2\pi \times 0.2$~MHz to $2\pi \times 2.4$~MHz. The reconstructed results, shown as purple squares in Fig.~\ref{fig:2}~(a), exhibit a dominant response at $2\pi \times 0.3$~MHz, while effectively suppressing responses at other frequencies. For comparison, we also implemented the conventional XY sequence with the same scanning frequency of $2\pi \times 0.3$~MHz. In contrast, the conventional XY sequence~(red diamonds), implemented with the same $\omega_{\text{scan}}$, displays additional responses at higher harmonics, indicating limited suppression of higher-order spectral components. For GD$_\perp$, the detuning $\Delta_s$ is swept over the same frequency range, and the reconstructed $|f(\omega)|$ is shown as purple squares in Fig.~\ref{fig:2}~(b). Compared with the CPMG sequence~(red diamonds), GD$_\perp$ similarly exhibits strong suppression of higher-order frequency components. Details of the reconstruction procedure for $|f(\omega)|$ and the implementation of XY and CPMG controls are provided in the Supplemental Material.

Compared with the conventional XY and CPMG sequences, the GD sequences exhibit a markedly enhanced resistance to higher-order noise. To explicitly demonstrate this advantage, we prepare the initial state $\ket{\psi_0}$, apply the GD sequences in the presence of higher-order noise, and measure the resulting fidelity of final state with respect to $\ket{\psi_0}$. For GD$_\parallel$ and XY, we set $\omega_\text{scan}=2\pi\times0.3$~MHz and the noise frequencies at $\omega_{n}=k\omega_\text{scan}$ with $k=3, 5$ and 7, respectively. The noisy amplitude  $b_n^\parallel$ is varied from 0 to $2\pi\times85$~kHz. As shown in Fig.~\ref{fig:2}~(c), both GD$_\parallel$ and XY are able to protect the initial state without noise~($b_n^\parallel=0$). With increasing $b_n^\parallel$, the final-state fidelity remains above 0.9 under GD$_\parallel$ control, whereas it decreases rapidly for the XY sequence. Moreover, for a fixed noise amplitude, lower-order harmonics (smaller $k$) induce a stronger degradation of fidelity, consistent with the reconstructed $|f(\omega)|$ shown in Fig.~\ref{fig:2}~(c). For the GD$\perp$ and CPMG sequences, we similarly set the noise detuning to $\Delta_n=k\omega_\text{scan}$ with $k=3, 5$ and 7, and the corresponding results are shown in Fig.~\ref{fig:2}~(d). Analogous to the GD$_\parallel$ case, GD$_\perp$ effectively protects the sensing state against higher-order noise, whereas the CPMG sequence exhibits a pronounced degradation of fidelity.

\begin{figure}[ht!bp]
\includegraphics[width=\linewidth]{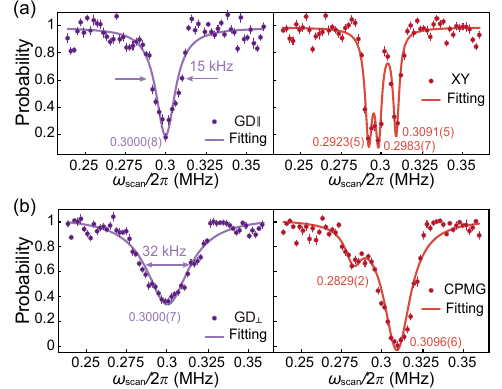}
\caption{The experimental results of frequency sensing with  
(a) the GD$_\parallel$ scheme~(purple dots) compared with the conventional XY scheme~(red dots), and (b) the GD$_\perp$ scheme~(purple dots) compared with the CPMG scheme~(red dots). Solid lines are fits to a Lorentzian function $P=A/(1 + (\omega_{\text{scan}} - \omega_s(\Delta_s))^2/\gamma^2)$, where 2$\gamma$ is the full width at half maximum, and $A$ is the signal amplitude.\label{fig:3}}
\end{figure}

Owing to their strong spectral selectivity and robustness, the GD sequences enable accurate frequency estimation even in the presence of higher-order noise components. For GD$_\parallel$ and XY schemes, we consider the parallel component $b_\parallel(t)$ in Eq.~\ref{Eq:ACparallel}, with a target signal at $\omega_s = 2\pi \times 0.3$~MHz and two noise components at $\omega_{n,1} = 2\pi \times 0.903$~MHz and $\omega_{n,2} = 2\pi \times 1.497$~MHz. The corresponding amplitudes are set to $b_s^\parallel =2\pi \times 24$~kHz and $b_{n,1}^\parallel = b_{n,2}^\parallel =2\pi \times 48$~kHz, respectively. Frequency estimation is performed by sweeping the scan frequency $\omega_\text{scan}$ over the range $2\pi \times [0.24, 0.36]$~MHz and recording the projection probability at each value. As shown in Fig.~\ref{fig:3}(a), the presence of noise induces three resonance dips in the XY scheme~(red dots), with biases $2\pi \times 8.7$~kHz, $2\pi \times 1.7$~kHz and $2\pi \times 9.1$~kHz relative to the target frequency, thereby preventing unambiguous identification of $\omega_s$. In contrast, noise-induced features are strongly suppressed under GD$_\parallel$ control, and only a single resonance dip at $2\pi \times 0.3$~MHz is observed~(purple dots). For GD$_\perp$ and CPMG sequences, we consider the perpendicular component $B_\perp(t)$ in Eq.~\ref{Eq:ACperp}, consisting of a target detuning $\Delta_s=2\pi \times 0.3$~MHz and a noisy detuning $\Delta_{n,1} = 2\pi \times 0.908$~MHz. The corresponding amplitudes are set as $b_s^\perp =2\pi \times 30$~kHz and $b_{n,1}^\perp =2\pi \times 60$~kHz, respectively. In both schemes, the scan frequency is swept over the same range $2\pi \times [0.24, 0.36]$~MHz. As shown in Fig.~\ref{fig:3}~(b), the noise induces substantial estimation biases of $2\pi \times 17.1$~kHz and $2\pi \times 9.6$~kHz in the CPMG scheme~(red dots). By contrast, the GD$_\perp$ protocol exhibits only a single resonance dip at $2\pi \times 0.3$~MHz~(purple dots).

\begin{figure}[htbp]
\includegraphics[width=\linewidth]{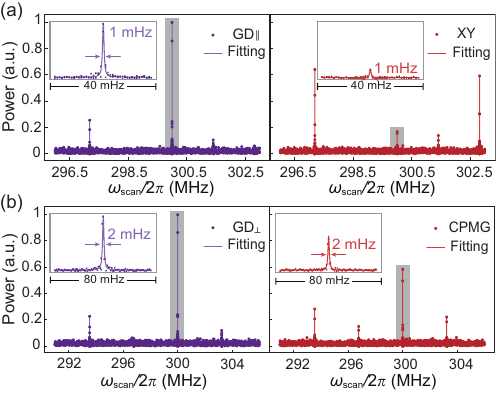}
\caption{The Fourier transform spectra of the signal-sensing experiment in the MHz (a) and GHz (b) regime. The insets show the fitting results of experimental spectrum slices in the shaded areas. \label{fig:4}}
\end{figure}

As shown in Fig.~\ref{fig:3}, the resolution of frequency estimation~(defined by the half-width of the full spectrum) obtained with GD$_\parallel$ and GD$_\perp$ are $2\pi\times 15$~kHz and $2\pi \times33$~kHz, respectively. The resolution can be further enhanced by incorporating the synchronized readout technique~\cite{Boss2017science,Schmitt2017science}. For the XY and GD$_\parallel$ schemes, $B_\parallel(t)$ is measured repeatedly with a repetition interval of $71~\mu$s, and a 23-minute time trace of photon counts is recorded.
For the CPMG and GD$_\perp$ schemes, $B_\perp(t)$ is measured repeatedly with a repetition interval of $31~\mu$s, and a 10-minute time trace is recorded. The corresponding Fourier spectra are shown in Fig.~\ref{fig:4}, tfrom which the frequency resolution is improved to 1~mHz for GD$_\parallel$ and 2~mHz for GD$_\perp$. Although synchronized readout also enhances the resolution in the XY and CPMG schemes, the presence of higher-order harmonic frequencies significantly degrades the signal-to-noise ratio as reflected in Fig.~\ref{fig:4}. In particular, in the XY scheme, multiple peaks appear in the measured spectrum, and the peak corresponding to the target signal (shaded area) is substantially lower than the spurious peaks induced by noise, rendering accurate frequency estimation challenging. More experimental details are provided in the Supplementary Materials.


In conclusion, we demonstrate accurate, high-resolution, and broadband detection of AC magnetic fields using a single NV center. By exploiting the single-frequency response of a quantum diamond sensor engineered through geodesic control, frequency-estimation bias is strongly suppressed even in multi-frequency noisy environments. The integration of geodesic sensing with heterodyne detection extends the accessible sensing bandwidth into the gigahertz regime, significantly broadening the scope of practical applications. Furthermore, the incorporation of synchronized readout enhances the frequency resolution to the millihertz level. Owing to its robustness, broadband operability, and compatibility with existing NV-based platforms, this approach is well suited for practical tasks such as nanoscale spectroscopy~\cite{Staudacher2013science, Nabeel2017science}, microwave-field characterization~\cite{Meinel2021NC, Wang2022SA}, and frequency-selective sensing in complex environments~\cite{Frey2017NC, Schmitt2021NPJqi}. Moreover, since the geodesic control can be implemented using standard control and readout techniques available in NV center platforms, our experimental demonstration indicates its compatibility with scalable sensing architectures, including multi-sensor~\cite{Jared2022science,Ji2024NP,Cheng2025PRX,Cambria2025PRX,Huxter2025PRL} and array-based implementations \cite{Shi2018NatureMethods,Cai2021RevSciInstrum,Wang2022SciAdv}.

\appendix
\begin{widetext}

\section{The conventional dynamical decoupling sensing}

For the MHz-regime signal, the Hamiltonian describing its projection parallel to the NV axis is $H_{\parallel }(t) =B_\parallel(t)\sigma_z/2$, with 
\begin{equation}
B_\parallel(t)=b_s^\parallel\cos(\omega_st + \varphi_s) + \sum_l b_{n,l}^\parallel \cos(\omega_{n,l} t + \varphi_{n,l}),
\end{equation}
where $b_s^\parallel$, $\omega_s$, and $\varphi_s$ denote the amplitude parallel to NV axis, frequency, and initial phase of the target signal, and $b_{n,l}^\parallel$, $\omega_{n,l}$, and $\varphi_{n,l}$ characterize the noise components.

The dynamical decoupling technique, such as XY sequence, can be implemented to extract the frequency information of the target signal while reducing the influence of noise signals. In the frequency sensing procedure, the quantum sensor is firstly prepared into the state $\ket{\psi_0^\parallel}=\ket{+} = (\ket{0} + \ket{1})/\sqrt{2}$. Then, the control pulse sequence consisting of $N_s$ pulse blocks is applied, given by
\begin{equation}
    \left(\frac{\tau}{2}-\pi_x-\tau-\pi_y-\tau-\pi_y-\tau-\pi_x-\frac{\tau}{2}\right )^{N_s}.
\end{equation}
 In one pulse block, $\pi_{x}$ and $\pi_{y}$ denote $\pi$ pulses implementing $\pi$ rotations about the $x$- and $y$-axes in Bloch sphere, respectively. $\tau$ is the free-evolution interval between adjacent $\pi$ pulses. For one pulse block, the center times of the $\pi_x$ pulses are at $\tau/2 + t_{\pi}/2$ and $7(\tau/2 + t_{\pi}/2)$, while those of the $\pi_y$ pulses are at $3(\tau/2 + t_{\pi}/2)$ and $5(\tau/2 + t_{\pi}/2)$, where $t_{\pi}$ denotes the pulse duration. The scan frequency is denoted as $\omega_{\text{scan}}=\pi/(\tau+t_\pi)$. An x-axis-polarized microwave (MW) field with resonant frequency $\omega_0$ is used to implemented this technique. In the laboratory frame, the control Hamiltonian driven by resonant MW field is
\begin{equation} 
H_{\text{XY}}^{\text{lab}}(t)=\Omega_{\text{XY}}(t)\cos\left (\omega_0 t+ \phi_{\text{XY}}(t)\right ) \sigma_x,
\end{equation}
where $\Omega_{\text{XY}}(t)$ and $\phi_{\text{XY}}(t)$ are the time-dependent Rabi frequency and phase of control field, respectively. In our experiment, during the pulse duration, the Rabi frequency is fixed at
$\Omega_{\text{XY}}(t) = \pi / t_{\pi} = 2\pi \times 10~\mathrm{MHz}$. The phase is set to $\phi_{\text{XY}}(t)=0$ for the $\pi_x$ pulse and to
$\phi_{\text{XY}}(t)=-\pi/2$ for the $\pi_y$ pulse. No control field is applied ($\Omega_{\text{XY}}(t) = 0$) between adjacent $\pi$ pulses. The pulse sequences is configured with $N_s = 8$. In the rotating frame defined by the unitary transformation $U_0 = \exp(-i\omega_0\sigma_z t / 2)$, after the rotating-wave approximation, the control Hamiltonian is
\begin{equation}
\begin{split}
        H_{\text{XY}}(t)&=U_0H_{\text{XY}}^{\text{lab}}(t)U_0^{\dagger}\\
    &=\frac{\Omega_{\text{XY}}(t)}{2}\left[\cos( \phi_{\text{XY}}(t))\sigma_x - \sin(\phi_{\text{XY}}(t))\sigma_y\right],\\
\end{split}
\end{equation}
the total Hamiltonian of the sensing can be written as
\begin{equation} 
    \begin{split}
    H_{\text{XY}}^{\text{tot}}(t)&=U_0[H_0+H_\parallel(t)+H_{\text{XY}}^{\text{lab}}(t)]U_0^{\dagger}-iU_0\frac{\mathrm{d} U_0^{\dagger}}{\mathrm{d} t} \\
    &=H_{\text{XY}}(t)+H_\parallel(t)\\
    &=\frac{\Omega_{\text{XY}}(t)}{2}\left[\cos( \phi_{\text{XY}}(t))\sigma_x - \sin(\phi_{\text{XY}}(t))\sigma_y\right]+B_\parallel(t)\frac{\sigma_z}{2}.
    \end{split}
\end{equation}
 In one pulse block, the $j$-th $\pi$ pulse ($\pi_x$ or $\pi_y$) is described by $U_j=\exp(-i\pi\sigma_x/2)$ or $U_j=\exp(-i\pi\sigma_y/2)$, respectively. After the $j$-th $\pi$ pulse, the evolution operator $U_{\text{XY}}$ corresponding to control Hamiltonian $H_{\text{XY}}(t)$ is given by
\begin{equation}\label{Eq:evolution_operator1_Appen}
    U_{\text{XY}}(j)=U_j\cdots U_1=\begin{cases}
 -i\sigma_x,~j=1\\i\sigma_z,~j=2
 \\i\sigma_x,~j=3
 \\I,~j=4
\end{cases} .
\end{equation}

Considering a time-dependent Hamiltonian $H(t)=H_s+H_c(t)$~\cite{Haeberlen1968PR}, where $H_s$ is the system Hamiltonian governing the system dynamics and $H_c(t)$ corresponds to the control field. The system evolution $U=\mathcal{T}\exp(-i \int_0^t H(t^\prime)dt^\prime)$ is complex to calculate, where $\mathcal{T}$ is time-ordering operator. To simplify the calculation of the system evolution, the system Hamiltonian is transformed into the interaction picture with respect to the control field, that is the system evolution is governed by
\begin{equation}
\widetilde{H}_s(t)=U_c^\dagger\, H_s \, U_c,
\end{equation}
where $U_c=\mathcal{T}\exp(-i \int_0^t H_c(t^\prime)dt^\prime)$. The system evolution can be calculated by $\mathcal{U}=\mathcal{T}\exp(-i \int_0^t \widetilde{H}_s(t^\prime)dt^\prime)$. Specifically, when the control field consists of $j$ pulses, $U_c$ can be rewritten as $U_c=U_j\cdots U_1$, $U_j$ is the unitary rotation corresponding to the $j$-th pulse, and this interaction-picture is also called by the toggling frame~\cite{Choi2020PRX}.

For analytical convenience, we transform the Hamiltonian 
$H_{\parallel}(t)$ into the toggling frame defined by the XY evolution operator $U_{\text{XY}}$, the effective sensing Hamiltonian is
\begin{equation} \label{Eq:effective1_Appen}
\begin{split}
    \widetilde{H}_{\parallel}^\prime(t)&=U_{\text{XY}}^\dagger H_{\parallel}(t) U_{\text{XY}}\\
    &=F_{\text{XY}}(t) B_\parallel(t)\frac{\sigma_z}{2}.
    \end{split}
\end{equation}
Under the instantaneous $\pi$-pulse approximation ($t_\pi \ll \tau$), the ideal modulation function of the dynamical decoupling sequence is a periodic square wave $F_{\text{XY}}(t)\in\{-1,1\}$, $F_{\text{XY}}(t)\approx\sum_{k=\text{odd}}^{\infty} 4\cos(k\omega_{\text{scan}}t)/k\pi$.

 Under the Hamiltonian $\widetilde{H}_{\parallel}^\prime(t)$ in Eq.~(\ref{Eq:effective1_Appen}), the prepared state undergoes a rotation about the $z$ axis, and the final state is $\ket{\psi_f}=\exp[-i\int_0^{t} \widetilde{H}_{\parallel}^\prime(t)dt^\prime]\ket{+}=[\ket{0} + \exp(i\Phi_{\text{XY}}(t))\ket{1}]/\sqrt{2}$. Assuming the initial phase of signal is $\varphi_s=0$, the accumulated phase $\Phi_{\text{XY}}(t)$ of final state is 

\begin{equation}\label{Phi_XY_Appen}
\begin{split}
    \Phi_{\text{XY}}(t)&=\int_0^tF_{\text{XY}}(t^\prime) B_\parallel(t^\prime)\,dt^\prime\\
    &=\int_0^t\sum_{k=\text{odd}}^{\infty} \frac{4}{k\pi}\cos\left(k\omega_{\text{scan}}t^\prime\right)\times[b_s^\parallel\cos(\omega_st^\prime ) + \sum_l b_{n,l}^\parallel\cos(\omega_{n,l} t^\prime + \varphi_{n,l})]\,dt^\prime\\
    &= \int_0^t \sum_k^{\infty} \Bigg\{\frac{2b_s^\parallel}{k\pi}\cos\left((\omega_s-k\omega_{\text{scan}})t^\prime\right)+ \sum_l \frac{2b_{n,l}^\parallel}{k\pi}\cos((\omega_{n,l}-k\omega_{\text{scan}})t^\prime + \varphi_{n,l}) +\\
    &\frac{2b_s^\parallel}{k\pi}\cos\left((k\omega_{\text{scan}}+\omega_s)t^\prime\right)+ \sum_l \frac{2b_{n,l}^\parallel}{k\pi}\cos((k\omega_{\text{scan}}+\omega_{n,l})t^\prime + \varphi_{n,l})\Bigg\}\,dt^\prime\\
    &\approx \int_0^t \sum_k^{\infty} \Bigg\{\frac{2b_s^\parallel}{k\pi}\cos\left((k\omega_{\text{scan}}-\omega_s)t^\prime\right)+ \sum_l \frac{2b_{n,l}^\parallel}{k\pi}\cos((k\omega_{\text{scan}}-\omega_{n,l})t^\prime + \varphi_{n,l}) \Bigg\}\,dt^\prime,\\
\end{split}
\end{equation}
where the high-frequency $(k\omega_{\text{scan}}+\omega_s)$ and $(k\omega_{\text{scan}}+\omega_{n,l})$ terms are neglected, and the conditions $|\Phi_{\text{XY}}(t)| < \pi$ is imposed. Finally, $\Phi_{\text{XY}}(t)$ is read out from the probability $P_{+}$ of projecting the final state onto $\ket{+}$
\begin{equation}
    P_{+}=|\braket{+|\psi_f}|^2=\frac{1}{2}+\frac{1}{2}\cos(\Phi_{\text{XY}}(t)).
\end{equation}

Under noise-free conditions ($b_{n,l}^\parallel=0$), the fundamental frequency component ($k=1$) of $F_{\text{XY}}(t)$ predicts that, as scan frequency is varied, the maximum accumulated phase is $\Phi_{\text{XY}}\approx b_sT/\pi$ when scan frequency matches the target-signal frequency $\omega_s$. The probability decreases the most. In this case, the frequency of the target signal is extracted. However, according to the higher odd-order terms of $F_{\text{XY}}(t)$, the sensor configured with scan frequency $\omega_{\text{scan}}$ can respond to signals with frequencies $k\omega_{\text{scan}}~(k = 3, 5, 7, \dots)$. When noise signals with frequencies $\omega_{n,l} \approx k\omega_{\text{scan}}~(k = 3, 5, 7, \dots)$ are present, these undesired responses would result in probability dips at frequencies away from the target frequency, leading to a biased estimation of the frequency.

\section{The conventional heterodyne technique}

For the fast-oscillating field, the Hamiltonian describing its projection perpendicular to the NV axis is $H_{\perp }(t) =B_\perp (t)\sigma_x$, with 
\begin{equation}
B_\perp (t)= b_s^\perp\cos(\omega_st + \varphi_s) + \sum_l b_{n,l}^\perp\cos(\omega_{n,l} t + \varphi_{n,l}),
\end{equation}
the first term represents the target signal, and the second term corresponds to the noise signal. After transforming this fast-oscillating field into the rotating frame with reference frequency $\omega_0$, we obtain a effective low-frequency field \cite{Meinel2021NC,Chu2021PRApplied}
\begin{equation} \label{signal2}
\begin{split}
H_{\perp}^{\prime}(t)
&=U_0H_{\perp}(t)U_0^{\dagger} \\
&= \frac{b_s^\perp}{2}[\cos(\Delta_s t+\varphi_s)\sigma_x-\sin(\Delta_s t+\varphi_s)\sigma_y]\\
&+\sum_l \frac{b_{n,l}^\perp}{2}[\cos(\Delta_{n,l} t + \varphi_{n,l})\sigma_x - \sin(\Delta_{n,l} t + \varphi_{n,l})\sigma_y],
\end{split}
\end{equation}
$\Delta_s = \omega_s - \omega_0$ and $\Delta_{n,l} = \omega_{n,l} - \omega_0$ denote the differences between the reference frequency and the signal frequencies. The amplitude of signals should satisfy $b_s \ll | \Delta_s |\ll\omega_s+\omega_0$ and $b_{n,l} \ll |\Delta_{n,l}|\ll\omega_{n,l}+\omega_0$. 

Generally, the CPMG sequence, i.e.,
\begin{equation}
    (\tau-\pi_{x}-\tau-\pi_{x})^{N_s},
\end{equation}
is used to measure the low-frequency detuning $\Delta_s$, from which the signal frequency $\omega_s=\omega_0+\Delta_s$ is extracted. 
In the frequency sensing procedure, the quantum sensor is firstly prepared into the state $\ket{0}$. Then, the control pulse sequence is applied.  In one pulse block, the center times of the $\pi_x$ pulses are at $\tau + t_{\pi}/2$ and $(2\tau + 3t_{\pi}/2)$. The scan frequency is denoted as $\omega_{\text{scan}}=\pi/(\tau+3t_\pi/4)$.
In the laboratory frame, the control Hamiltonian driven by resonant MW filed is
\begin{equation}
H_{\text{CPMG}}^{\text{lab}}(t)=\Omega_{\text{CP}}(t)\cos(\omega_0 t)\sigma_x.
\end{equation}
where $\Omega_{\text{CP}}(t)$ is the time-dependent Rabi frequency of control field. Within the pulse duration, the Rabi frequency is fixed at $\Omega_{\text{CP}}(t) = \pi / t_{\pi}=2\pi \times10$~MHz, the phase is set to zero. The pulse sequence is configured with $N_s = 4$. Between two  adjacent $\pi$ pulses, the control field is switched off $\Omega_{\text{CP}}(t) = 0$. In the rotating frame defined by the unitary transformation $U_0$, the control Hamiltonian is
\begin{equation}
H_{\text{CPMG}}(t)=U_0H_{\text{CPMG}}^{\text{lab}}(t)U_0^{\dagger}=\frac{\Omega_{\text{CP}}(t)}{2}\sigma_x.
\end{equation}
The total Hamiltonian during sensing is
\begin{equation}
    \begin{split}
    H_{\text{CPMG}}^{\text{tot}}(t)
    &=U_0[H_0+H_{\perp}(t)+H_{\text{CPMG}}^{\text{lab}}(t)]U_0^{\dagger}-iU_0\frac{\mathrm{d} U_0^{\dagger}}{\mathrm{d} t} \\
    &=H_{\text{CPMG}}(t)+H_{\perp}^{\prime}(t).
    \end{split}
\end{equation}
In the control sequence, after the $j$-th $\pi$ pulse ($\pi_x$), the evolution operator $U_{\text{CPMG}}$ corresponding to control Hamiltonian $H_{\text{CPMG}}$ can be expressed as
\begin{equation}\label{Eq:evolution_operator2_Appen}
    U_{\text{CPMG}}(j)=U_j\cdots U_1=
 e^{-\frac{\pi}{2}j}\cos^2(\frac{j\pi}{2})(\sigma_x+I).
\end{equation}

Similar to XY scheme, in the interacting picture defined by evolution operator $U_{\text{CPMG}}$, the effective sensing Hamiltonian is

\begin{equation}\label{Eq:effective3_Appen}
\begin{split}
    \widetilde{H}_{\perp}^\prime(t)
    &=U_{\text{CPMG}}H_\perp^\prime(t) U_{\text{CPMG}}^{\dagger} \\
    &= B_{X}(t)\frac{\sigma_x}{2}-F_{\text{CP}}(t) B_{Y}(t)\frac{\sigma_y}{2},\\
    \end{split}
\end{equation}
with
\begin{equation}
\begin{split}
B_{X}(t)&=b_s^\perp\cos(\Delta_st + \varphi_s) + \sum_l b_{n,l}^\perp\cos(\Delta_{n,l} t + \varphi_{n,l}),\\
     B_{Y}(t)&=b_s^\perp\sin(\Delta_st + \varphi_s) + \sum_l b_{n,l}^\perp\sin(\Delta_{n,l} t + \varphi_{n,l}).\\
\end{split}
\end{equation}
Under the instantaneous $\pi$-pulse approximation, $F_{\text{CP}}(t)\approx\sum_{k=\text{odd}}^{\infty } 4\sin(k \omega_{\text{scan}}t)/(k\pi)$ is the periodic modulation function. Under the condition $\Delta_s\gg b_s^\perp$, the effect of the first term in  $\widetilde{H}_{\perp}^\prime(t)$ can be neglected, and thus the effective Hamiltonian simplifies to $\widetilde{H}_{\perp}^\prime(t) \approx -F_{\text{CP}}(t)B_Y(t)\sigma_y/2$.

The electron spin evolve under Hamiltonian $\widetilde{H}_{\perp}^\prime(t)$ in Eq.~(\ref{Eq:effective3_Appen}), and the final state is $\ket{\psi_f}=\exp[-i\int_0^{t} \widetilde{H}_{\perp}^\prime(t^\prime)dt^\prime]\ket{0}=[\cos(\Phi_{\text{CP}}(t)/2)\ket{0} - \sin(\Phi_{\text{CP}}(t)/2)\ket{1}]/\sqrt{2}$. Assuming the initial phase of signal is $\varphi_s=0$, the rotation angle $\Phi_{\text{CP}}(t)$ of final state is 

\begin{equation}\label{Phi_CP_Appen}
\begin{split}
    \Phi_{\text{CP}}(t)&=\int_0^tF_{\text{CP}}(t^\prime) B_Y(t^\prime)\,dt^\prime\\
    &=\int_0^t\sum_{k=\text{odd}}^{\infty} \frac{4}{k\pi}\sin\left(k\omega_{\text{scan}}t^\prime\right)\times[b_s^\perp\sin(\Delta_st^\prime ) + \sum_l b_{n,l}^\perp\sin(\Delta_{n,l} t^\prime + \varphi_{n,l})]\,dt^\prime\\
    &= \int_0^t \sum_k^{\infty} \Bigg\{\frac{2b_s^\perp}{k\pi}\cos\left((\Delta_s-k\omega_{\text{scan}})t^\prime\right)+ \sum_l \frac{2b_{n,l}^\perp}{k\pi}\cos((\Delta_{n,l}-k\omega_{\text{scan}})t^\prime + \varphi_{n,l})-\\
    &\frac{2b_s^\perp}{k\pi}\cos\left((k\omega_{\text{scan}}+\Delta_s)t^\prime\right)- \sum_l \frac{2b_{n,l}^\perp}{k\pi}\cos((k\omega_{\text{scan}}+\Delta_{n,l})t^\prime + \varphi_{n,l})\Bigg\}\,dt^\prime\\
    &\approx \int_0^t \sum_k^{\infty} \Bigg\{\frac{2b_s^\perp}{k\pi}\cos\left((k\omega_{\text{scan}}-\Delta_s)t^\prime\right)+ \sum_l \frac{2b_{n,l}^\perp}{k\pi}\cos((k\omega_{\text{scan}}-\Delta_{n,l})t^\prime + \varphi_{n,l}) \Bigg\}\,dt^\prime,\\
\end{split}
\end{equation}

where we let $\Phi_{\text{CP}}(t)$ to satisfy $|\Phi_{\text{CP}}(t)| < \pi$.
Finally, the probability of projecting the final state onto $\ket{0}$ is, $P_{\ket{0}}=|\braket{0|\psi_f}|^2=\frac{1}{2}[1+\cos(\Phi_{\text{CP}}(t))]$.
Under noise-free conditions, the probability decreases the most when $\tau$ is tuned such that scan frequency $\omega_{\text{scan}}$ matches $\Delta_s$, the frequency of the target signal is extracted. However, noise contributions from higher odd-harmonic frequencies can lead to inaccurate frequency estimation.

\section{The geodesic sensing protocol} 
During the frequency-sensing procedure in the MHz regime, the quantum sensor is firstly prepared into the state $\ket{\psi_0^\parallel}=\ket{+}=(\ket{0}+\ket{1})/\sqrt{2}$. Then, the pulse sequence of the GD$_\parallel$ scheme is applied. The geodesic pulse sequence in GD$_\parallel$ scheme is 
\begin{equation}
    {\left(\frac{\tau_{\text{GD}}}{2}-\pi_{\phi_1}-\tau_{\text{GD}} -\pi_{\phi_2}-\cdots-\tau_{\text{GD}} -\pi_{\phi_N}-\frac{\tau_{\text{GD}}}{2}\right)}^{N_s},
\end{equation}
 where a total of $N$ $\pi$ pulses are applied within one period $T_{\text{scan}} = N(\tau_{\text{GD}}+t_\pi)$, $\tau_{\text{GD}}$ is the time interval between adjacent $\pi$ pulses. The central moment of the $j$-th $\pi$ pulse is given by $T_j = T_{\text{scan}}[(2j - 1)/2N]$, where $j = 1, 2, \dots, N$ labels the pulse index within one period. $\pi_{\phi_j}$ denotes a $\pi$ rotation about the axis $(\sin\phi_j, 0, -\cos\phi_j)$, where $\phi_j=2\pi T_j/T_{\text{scan}}$. Experimentally, the geodesic pulse sequence is implemented using a microwave~(MW) field described in the laboratory frame by
 \begin{equation}
     \begin{split}
         H_{\text{GD}_\parallel}^{\text{lab}}(t)=&[\Omega_{\text{GD}}(t)\sin \phi_{\text{GD}}(t)] \times\\
    &\cos\left[\omega_0t-\int_{0}^{t}\Omega_{\text{GD}}(t^\prime) \cos \phi_{\text{GD}}(t^\prime)dt^\prime\right]\sigma_x,
     \end{split}
 \end{equation}
where $\Omega_{\text{GD}}(t) \sin \phi(t)$ denotes the instantaneous Rabi frequency, while $\omega_0$ and $\int_{0}^{t}\Omega_{\text{GD}}(t^\prime) \cos \phi_{\text{GD}}(t^\prime)d t^\prime$ define the MW frequency and phase, respectively. During each pulse interval $t \in [T_j-t_\pi/2, T_j+t_\pi/2]$, where $t_\pi$ denotes the pulse duration, the Rabi parameter is fixed at $\Omega_{\text{GD}}(t) = \pi / t_{\pi}=2\pi \times 10$~MHz, and the phase parameter is fixed at $\phi_{\text{GD}}(t)=\phi_j=2\pi T_j/T_{\text{scan}}$. No control field is applied ($\Omega_{\text{GD}}(t) =0$) between adjacent $\pi$ pulses. The pulse sequence is configured with $N_s = 8$ and $N = 10$.
In the rotating frame defined by $U = \exp\{-i[\omega_0 t - \int_{0}^{t} \Omega_{\text{GD}}(t^\prime) \cos \phi_{\text{GD}}(t^\prime) dt^\prime] \sigma_z / 2\}$, the control Hamiltonian is given by
\begin{equation}
\begin{split}
    H_{\text{GD}_\parallel}(t)&=UH_{\text{GD}_\parallel}^{\text{lab}}(t)U^{\dagger} \\
    &=\frac{\Omega_{\text{GD}}(t)}{2}(\sin\phi_{\text{GD}}(t)\sigma_x-\cos\phi_{\text{GD}}(t)\sigma_z)\\
    &=\frac{\Omega_{\text{GD}}(t)}{2}(\ket{+_{\phi}}\bra{+_{\phi}}-\ket{-_{\phi}}\bra{-_{\phi}}).\
    \end{split}
\end{equation}
When the number of pulses $N$ is sufficiently large, the evolution path under control field are 
$\ket{+_{\phi}}=\cos(\phi_{\text{GD}}(t)/2)\ket{1}+\sin(\phi_{\text{GD}}(t)/2)\ket{0}$ and $\ket{-_\phi}=-\sin(\phi_{\text{GD}}/2)\ket{1}+\cos(\phi_{\text{GD}}/2)\ket{0}$. Considering the AC magnetic signal field $B_\parallel(t)$, the total Hamiltonian of the sensing in the rotating frame is
\begin{equation}
\begin{split}
    H_{\text{GD}_\parallel}^{\text{tot}}(t)&=U[H_0+H_{\parallel}(t)+H_{\text{GD}_\parallel}^{\text{lab}}(t)]U^{\dagger}-iU\frac{\mathrm{d} U^{\dagger}}{\mathrm{d} t} \\
    &=H_{\text{GD}_\parallel}(t)+H_\parallel(t)\\
    &= \frac{\Omega_{\text{GD}}(t)}{2}(\ket{+_{\phi}}\bra{+_{\phi}}-\ket{-_{\phi}}\bra{-_{\phi}})+\frac{B_\parallel(t)}{2}\sigma_z.
    \end{split}
\end{equation}

In one pulse block, after $j$-th $\pi$ pulse described by $U_j=\exp[-i\pi(\sin\phi_j\sigma_x-\cos\phi_j\sigma_z)/2]$, the evolution operator $U_{\text{GD}_\parallel}$ is given by 
\begin{equation}\label{Eq:evolution_operator3_Appen}
\begin{split}
    U_{\text{GD}_\parallel}(j)&=U_jU_{j-1}\cdots U_2 U_1\\
    &=e^{-i\varphi_+^\parallel}\ket{+_{\bar{\phi}_j}^{\parallel}}\bra{1}+e^{-i\varphi_-^\parallel}\ket{-_{\bar{\phi}_j}^{\parallel}}\bra{0},
\end{split}
\end{equation}
where $\varphi_{\pm}^\parallel=\pm j\pi/2$, $\ket{+_{\bar{\phi}_j}^{\parallel}}=\cos(\bar{\phi}_j/2)\ket{1}+\sin(\bar{\phi}_j/2)\ket{0}$ and $\ket{-_{\bar{\phi}_j}^{\parallel}}=-\sin(\bar{\phi}_j/2)\ket{1}+\cos(\bar{\phi}_j/2)\ket{0}$, with $\bar{\phi}_j=2\pi j/N$. We obtain $U_{\text{GD}_\parallel}^\dagger(j)\sigma_zU_{\text{GD}_\parallel}(j)\approx\cos\bar{\phi}_j\sigma_z$, where we neglect the fast varying terms.
We transform the Hamiltonian 
$H_{\parallel}$ into the interacting picture defined by the evolution operator $U_{\text{GD}_\parallel}$, the effective sensing Hamiltonian is
\begin{equation} \label{Eq:effective2_Appen}
\begin{split}
    \widetilde{H}_\parallel(t)&=U_{\text{GD}_\parallel}^\dagger H_{\parallel}(t) U_{\text{GD}_\parallel}\\
    &=F_{\text{GD}_\parallel}(t)B_\parallel(t)\frac{\sigma_z}{2}.
    \end{split}
\end{equation}
Under the instantaneous $\pi$-pulse approximation, the ideal modulation function in GD$_\parallel$ approaches a single-frequency modulation function in the limit of sufficiently large $N$, $F_{\text{GD}_\parallel}(t)\approx \cos(2\pi t /T_{\text{scan}})$. The scan frequency is $\omega_{\text{scan}}=2\pi/T_{\text{scan}}$.
 The electron spin evolve under Hamiltonian $\widetilde{H}_{\parallel}(t)$ in Eq.~(\ref{Eq:effective2_Appen}), and the final state is $\ket{\psi_f^\parallel}=[\ket{0} + \exp(i\Phi_{\text{GD}_\parallel}(t))\ket{1}]/\sqrt{2}$, where $\Phi_{\text{GD}_\parallel}(t)$ the accumulated phase of prepared state. Assuming the initial phase of signal is $\varphi_s=0$, the accumulated phase is 

\begin{equation}\label{Phi_GDy_Appen}
\begin{split}
    \Phi_{\text{GD}_\parallel}(t)
    &=\int_0^tF_{\text{GD}_\parallel}(t^\prime) B_\parallel(t^\prime)\,dt^\prime\\&=\int_0^t\cos(\omega_{\text{scan}}t^\prime) B_\parallel(t^\prime)\,dt^\prime\\
    &=\int_0^t \Bigg\{\frac{b_s^\parallel}{2}\cos((\omega_s-\omega_{\text{scan}})t^\prime)+ \sum_l \frac{b_{n,l}^\parallel}{2}\cos((\omega_{n,l}-\omega_{\text{scan}}) t^\prime + \varphi_{n,l})+\\
    &\frac{b_s^\parallel}{2}\cos((\omega_s+\omega_{\text{scan}})t^\prime)+ \sum_l \frac{b_{n,l}^\parallel}{2}\cos((\omega_{n,l}+\omega_{\text{scan}}) t^\prime + \varphi_{n,l})\Bigg\}\,dt^\prime\\
    &\approx\int_0^t \Bigg\{\frac{b_s^\parallel}{2}\cos((\omega_s-\omega_{\text{scan}})t^\prime)+ \sum_l \frac{b_{n,l}^\parallel}{2}\cos((\omega_{n,l}-\omega_{\text{scan}}) t^\prime + \varphi_{n,l})\Bigg\}\,dt^\prime,\\
\end{split}
\end{equation}

where the conditions $|\Phi_{\text{GD}_\parallel}(t)| < \pi$ is imposed. $\Phi_{\text{GD}_\parallel}(t)\neq0$ when  $\omega_{\text{scan}}=\omega_s$, whereas $\Phi_{\text{GD}_\parallel}(t)\approx0$ for $\omega_{\text{scan}}\neq\omega_s$. Finally, the probability of projecting the final state onto $\ket{+}$ is
\begin{equation}
    P_{\text{+}}=\frac{1}{2}+\frac{1}{2}\cos(\Phi_{\text{GD}_\parallel}(t)).
\end{equation}
 Consequently, the probability decreases most when $\omega_{\text{scan}}$ is tuned to the target frequency $\omega_s$, noise contributions at other frequencies are strongly suppressed, enabling high-accuracy determination of the target signal frequency.

For the robust heterodyne sensing (GD$_\perp$ scheme), the prepared state is $\ket{\psi_0^\perp}=\ket{L}=(\ket{0}+i\ket{1})/\sqrt{2}$, then the geodesic control sequence of the GD$_\perp$ scheme can be employed measure low-frequency $\Delta_s$. The geodesic control sequence takes the same form as that in GD$_\parallel$ scheme. $\pi_{\phi_j}$ is a $\pi$ rotation about the axis $(\cos\phi_j, -\sin\phi_j, 0)$. The control Hamiltonian driven by MW field in the laboratory frame is $H_{\text{GD}_\perp}^{\text{lab}}(t)=\Omega_{\text{GD}}(t)\cos[\omega_0 t+\phi_{\text{GD}}(t)]\sigma_x$, the parameters $\Omega_{\text{GD}}(t)$ and $\phi_{\text{GD}}(t)$ are the same as those in GD$_\parallel$ scheme. In the rotating frame defined by the unitary transformation $U_0$, the control Hamiltonian is
\begin{equation}
\begin{split}
    H_{\text{GD}_\perp}(t)&=U_0H_{\text{GD}_\perp}^{\text{lab}}(t)U_0^{\dagger} \\
    &=\frac{\Omega_{\text{GD}}(t)}{2} (\cos \phi_{\text{GD}}(t) \sigma_x-\sin \phi_{\text{GD}}(t) \sigma_y)\\
    &=\frac{\Omega_{\text{GD}}(t)}{2}(\ket{+_{\phi^\prime}}\bra{+_{\phi^\prime}}-\ket{-_{\phi^\prime}}\bra{-_{\phi^\prime}}),\\
    \end{split}
\end{equation}
the evolution path under control field is $\ket{\pm_{\phi^\prime}}=(\ket{1}\pm\exp(i\phi_{\text{GD}}(t))\ket{0})/\sqrt{2}$. 
Considering the AC signal field $B_\perp(t)$, the total Hamiltonian during sensing is
\begin{equation}
\begin{split}
    H_{\text{GD}_\perp}^{\text{tot}}(t)
    &=U_0[H_0+H_{\perp}(t)+H_{\text{GD}_\perp}^{\text{lab}}(t)]U_0^{\dagger}-iU_0\frac{\mathrm{d} U_0^{\dagger}}{\mathrm{d} t} \\
    &=H_{\text{GD}_\perp}(t)+H_\perp^\prime(t).\\
    \end{split}
\end{equation}

In one pulse block, after $j$-th $\pi$ pulse described by $U_j^\prime=\exp[-i\pi(\cos\phi_j\sigma_x-\sin\phi_j\sigma_y)/2]$, the evolution operator $U_{\text{GD}_\perp}$ is given by 
\begin{equation}\label{Eq:evolution_operator4_Appen}
\begin{split}
    U_{\text{GD}_\perp}(j)&=U_j^\prime U_{j-1}^\prime \cdots U_2^\prime U_1^\prime\\
    &=e^{-i\varphi_+^\perp}\ket{+_{\bar{\phi}_j}^{\perp}}\bra{+}+e^{-i\varphi_-^\perp}\ket{-_{\bar{\phi}_j}^{\perp}}\bra{-},
\end{split}
\end{equation}
where $\varphi_{+}^\perp=j\pi(1+1/N)$, $\varphi_{-}^\perp=j\pi/N$,  $\ket{+_{\bar{\phi}_j}^{\perp}}=(\ket{1}+\exp(i\bar{\phi}_j)\ket{0})/\sqrt{2}$ and $\ket{-_{\bar{\phi}_j}^{\perp}}=(\ket{1}-\exp(i\bar{\phi}_j)\ket{0})/\sqrt{2}$, with $\bar{\phi}_j=2\pi j/N$. We obtain $U_{\text{GD}_\perp}^\dagger(j)\sigma_xU_{\text{GD}_\perp}(j)\approx\cos\bar{\phi}_j\sigma_x$ and $U_{\text{GD}_\perp}^\dagger(j)\sigma_yU_{\text{GD}_\perp}(j)\approx-\sin\bar{\phi}_j\sigma_x$, where we neglect the fast varying terms. 
In the interacting picture defined by control Hamiltonian $U_{\text{GD}_\perp}$, the effective sensing Hamiltonian is
\begin{equation} \label{Eq:effective4_Appen}
\begin{split}
    \widetilde{H}_{\perp}(t)&=U_{\text{GD}_\perp}^\dagger H_{\perp}^\prime(t) U_{\text{GD}_\perp}\\
    &\approx\cos\left(\omega_{\text{scan}}t\right)B_X(t)\frac{\sigma_x}{2}+\sin\left(\omega_{\text{scan}}t\right)B_Y(t)\frac{\sigma_x}{2}\\
    &= \frac{b_s^\perp}{2}\cos\left((\Delta_s-\omega_{\text{scan}})t+\varphi_s\right)\sigma_x+ \sum_l \frac{b_{n,l}^\perp}{2}\cos((\Delta_{n,l}-\omega_{\text{scan}})t + \varphi_{n,l})\sigma_x\\
    &\approx\cos\left(\omega_{\text{scan}}t\right)B_X(t)\sigma_x\\
    &=F_{\text{GD}_\perp}(t)B_X(t)\sigma_x.\\
\end{split}
\end{equation}
i.e., when the sensing duration $T_s$ satisfies 
$\omega_{\text{scan}}+\Delta_s \gg 1/T_s$ and 
$\omega_{\text{scan}}+\Delta_{n,l} \gg 1/T_s$, 
the modulation function in the GD$_\perp$ scheme can be approximated as
$F_{\text{GD}_\perp}(t) \approx \cos(\omega_{\text{scan}} t)$. The effective Hamiltonian implements a rotation about the $x$-axis acting on the prepared state. The effective Hamiltonian implements a rotation along x-axis on prepared state, the final state is $\ket{\psi_f^\perp}=\cos(\Phi_{\text{GD}_\perp}(t)/2)\ket{L}+\sin(\Phi_{\text{GD}_\perp}(t)/2)\ket{R}$, with the rotation angle 

\begin{equation}\label{Phi_GDz_Appen}
\begin{split}
    \Phi_{\text{GD}_\perp}(t)
    &=\int_0^tF_{\text{GD}_\perp}(t^\prime) B_X(t^\prime)\,dt^\prime\\&=\int_0^t\cos(\omega_{\text{scan}}t^\prime) B_X(t^\prime)\,dt^\prime\\
    & =\int_0^t \Bigg\{b_s^\perp\cos((-\omega_{\text{scan}}+\Delta_s)t^\prime)+\sum_l b_{n,l}^\perp\cos((-\omega_{\text{scan}}+\Delta_{n,l}) t^\prime + \varphi_{n,l})+\\
    &b_s^\perp\cos((\omega_{\text{scan}}+\Delta_s)t^\prime)+\sum_l b_{n,l}^\perp\cos((\omega_{\text{scan}}+\Delta_{n,l}) t^\prime + \varphi_{n,l})\Bigg\}\,dt^\prime\\
    & \approx\int_0^t \Bigg\{b_s^\perp\cos((-\omega_{\text{scan}}+\Delta_s)t^\prime)+\sum_l b_{n,l}^\perp\cos((-\omega_{\text{scan}}+\Delta_{n,l}) t^\prime + \varphi_{n,l})\Bigg\}\,dt^\prime.\\
\end{split}
\end{equation}

When scan frequency $\omega_{\text{scan}}$ satisfies $\omega_{\text{scan}}=\Delta_s=\omega_s-\omega_0$, the rotation angle reaches its maximum, the probability of projecting the final state onto $\ket{L}$, $P_L=|\braket{L|\psi_f}|^2=\frac{1}{2}[1+\cos(\Phi_{\text{GD}_\perp}(t)]$, reaches its minimum. Consequently, the target frequency $\omega_s = \omega_0 + \Delta_s$ can be obtained by accurately measuring the slow frequency $\Delta_s$.

\section{The Fourier components of the modulation function}

In the MHz-regime reconstruction experiment, we consider a single random magnetic signal $B_R(t)$ with a projection parallel to the NV axis
\begin{equation}
    \beta_\parallel(t)=b_R^\parallel \cos(\omega_R t+\varphi_R ),
\end{equation}
 where $b_R^\parallel$, $\omega_R$ and $\varphi_R$ are the amplitude along the NV axis, frequency, and random phase of the signal, respectively. We prepare the spin state in $\ket{+}$ and then apply the control sequence of the XY or GD$_\parallel$ scheme. In the interacting picture defined by evolution operator in Eqs.~(\ref{Eq:evolution_operator1_Appen}) and (\ref{Eq:evolution_operator3_Appen}), the effective sensing Hamiltonian of XY and GD$_\parallel$ schemes is
\begin{equation}\label{Eq:effective_Appen1}
\begin{split}
    \widetilde{H}_R^\parallel(t) 
    &=F_{z}(t)\beta_\parallel(t) \sigma_{z}/2,\\
    \end{split}
\end{equation}
$F_z(t)$ is the modulation function for the signal. The accumulated phase after the frequency-sensing procedure governed by Eq.~(\ref{Eq:effective_Appen1})  is
\begin{equation} 
\Phi_\parallel(t)=\int^{t}_0F_z(t^{\prime})\beta_\parallel(t^{\prime})dt^{\prime}.
\end{equation}
The phase $\Phi_\parallel(t)$ is read out from the probability of projecting final state onto $\ket{+}$
\begin{equation}
    \Phi_\parallel(t)=\frac{\pi}{2}-\arcsin(2P_+-1).
\end{equation}
We repeat the above procedure with the same scan frequency, varying only the phase $\varphi_R$. The phase $\varphi_R$ remain constant within each frequency sensing procedure but varies randomly between different procedures \cite{Suter2016RMPhys}, and the accumulated phase $\Phi_\parallel(t)$ thus varies randomly between procedures. Take the average over all accumulated phases
\begin{equation}\label{Eq_Phi_average}
\overline{\Phi_\parallel^2(t)}=\int^{t}_0 \int^{t}_0 F_z(t^{\prime})F_z(t^{\prime \prime})\overline{\beta_\parallel(t^{\prime})\beta_\parallel(t^{\prime\prime})}dt^{\prime}dt^{\prime\prime},
\end{equation}
where the overbar represents the ensemble average. 

The power spectral density (PSD) $S(\omega)=(1/\sqrt{2 \pi})\int^{\infty}_{-\infty}g(\Delta t)e^{-i\omega\Delta t}d\Delta t$ is obtained from the Fourier transformation of two-time correlation function $g(\Delta t)=\overline{\beta_\parallel(t^{\prime})\beta_\parallel(t^{\prime}+\Delta t)}$, $\Delta t$ is defined as the relative time interval between $t^{\prime}$ and $t^{\prime\prime}$. The Fourier component $|f(\omega,t)|$ of modulation function $F_z(t)$ is $|f(\omega,t)|=|(1/\sqrt{2 \pi})\int^{t}_{0}F_z(t^{\prime})e^{-i\omega t^{\prime}}dt^{\prime}|$. Base on the filter function $|f(\omega,t)|^2$ and the PSD function $S(\omega)$, Eq.~(\ref{Eq_Phi_average}) is converted into
\begin{equation} \label{phi_average}
\overline{\Phi_\parallel^2(t)}=\sqrt{2\pi}\int^{\infty}_{0}  |f(\omega,t)|^2 S(\omega)d\omega. 
\end{equation}

By choosing the signal phase $\varphi_R$ to be uniformly distributed in $[0,2\pi]$, $\beta_\parallel(t)$ becomes a wide-sense stationary (WSS) signal, so the time average $\langle \beta_\parallel(t^{\prime})\beta_\parallel(t^{\prime}+\Delta t) \rangle_{t^{\prime}}$ is equivalent to the ensemble average $\overline{\beta_\parallel(t^{\prime})\beta_\parallel(t^{\prime}+\Delta t)}$, where $\left \langle \dots \right \rangle_{t^\prime} $ denotes the average over all times $t^{\prime}$. This WSS signal satisfies
\begin{equation}
    \begin{split}
    &\overline{\beta_\parallel(t^{\prime})\beta_\parallel(t^{\prime}+\Delta t)}\\
    &=\frac{(b_R^\parallel)^2}{2\pi}\int_0^{2\pi}\cos(\omega_Rt^{\prime}+\theta)\cos(\omega_R(t^{\prime}+\Delta t)+\theta)d\theta\\
    &=\frac{(b_R^\parallel)^2}{4\pi}\int_0^{2\pi}[\cos(\omega_R(2t^{\prime}+\Delta t)+2\theta)+\cos(\omega_R\Delta t)]d\theta\\
    &=\frac{(b_R^\parallel)^2}{2}\cos(\omega_R \Delta t).\\
    \end{split}
\end{equation}
$S(\omega)$ can be written as
\begin{equation} \label{Eq:PSD}
\begin{split}
S(\omega)&=\frac{\sqrt{2\pi}(b_R^\parallel)^2}{4}  \delta(\omega-\omega_R) \\
& \propto \delta(\omega-\omega_R).
\end{split}
\end{equation}
Therefore, by averaging over all accumulated phases obtained with the random signal and substituting Eq.~(\ref{Eq:PSD}) into Eq.~(\ref{phi_average}), we obtain $|f(\omega_R,t)|$
\begin{equation}
|f(\omega_R,t)|^2=\frac{2\overline{\Phi_\parallel^2(t)}}{\pi(b_R^\parallel)^2}. 
\end{equation}

 We reconstruct the Fourier components of the modulation function, taking into account the finite-width control pulses ($t_\pi=50$~ns) in XY and GD$_\parallel$ experiments. The scan frequency is $2\pi \times 0.3$~MHz. The pulse sequences are configured with $N_s = 8$ and $N = 10$. To equivalently simulate a WSS signal, in each procedure the signal phase $\varphi_R$ is chosen uniformly at random from the set $\{0, \pi/3, 2\pi/3, \dots, 2\pi\}$ \cite{Frey2017NC,Frey2020PRApplied}. To reconstruct $|f(\omega,T)|$, we vary the frequency $\omega_R$ of the random signal within four separate frequency ranges $\omega_R\in2\pi\times[0.17,0.56]$~MHz, $\omega_R\in2\pi\times[0.64,1.24]$~MHz, $2\pi\times[1.32,1.76]$~MHz and $2\pi\times[1.85,2.36]$~MHz. The amplitude $b_R$ are $2\pi\times24$~kHz, $2\pi\times72$~kHz, $2\pi\times120$~kHz and $2\pi\times168$~kHz for the four ranges, respectively.

In the GHz-regime experiment, the single random magnetic signal has a projection perpendicular to the NV axis
\begin{equation}
    \beta_\perp(t)=b_R^\perp \cos(\omega_R t+\varphi_R ),
\end{equation}
transforming this fast-oscillating field into the rotating frame with reference frequency $\omega_0$
\begin{equation} 
\begin{split}
H_{R}^{\perp}(t)
&=U_0(\beta_\perp(t)\sigma_x)U_0^{\dagger}\\
&= \frac{b_R^\perp}{2}[\cos(\Delta_R t+\varphi_R)\sigma_x-\sin(\Delta_R t+\varphi_R)\sigma_y],\\
\end{split}
\end{equation}
$\Delta_R = \omega_R - \omega_0$ denotes the differences between the reference frequency and the signal frequency. We prepare the spin state in $\ket{0}$ for the CPMG sequence and in $\ket{L}$ for the GD$_\perp$ scheme, then we apply control sequence of CPMG or GD$_\perp$.
 In the interacting picture defined by evolution operator in Eqs.~(\ref{Eq:evolution_operator2_Appen}) and (\ref{Eq:evolution_operator4_Appen}), the effective sensing Hamiltonian is
\begin{subequations}\label{Eq:effective_Appen2}
    \begin{align}
    \widetilde{H}_R^\perp(t)
    &\approx -\frac{1}{2}F_{\text{CP}}(t)\sin(\Delta_R t+\varphi_R)\sigma_y,~\text{for~CPMG},\\
    \widetilde{H}_R^\perp(t) 
    &\approx F_{\text{GD}_\perp}(t)\cos(\Delta_R t+\varphi_R)\sigma_x,~\text{for~GD}_\perp.
    \end{align}
\end{subequations}
 The rotation angle of prepared state during the frequency-sensing procedure is
\begin{subequations}
\begin{align}
    \Phi_\perp(t)&=\int^{t}_0F_{\text{CP}}(t^\prime)\sin(\Delta_R t^\prime+\varphi_R)dt^{\prime},~\text{for CPMG},\\
    \Phi_\perp(t)&=\int^{t}_0 2F_{\text{GD}_\perp}(t^\prime)\cos(\Delta_R t^\prime+\varphi_R)dt^{\prime},~\text{for GD}_\perp.
\end{align}
\end{subequations}
Finally, $\Phi_\perp(t)$ is read out from the probability of projecting final state onto initial state. The above procedure is repeated with the same scan frequency ($2\pi \times 0.3$~MHz) and pulse duration ($50$~ns). The pulse sequence are configured with $N_s = 4$ and $N = 10$. The signal phase $\varphi_R$ is chosen uniformly at random from the set $\{0, \pi/3, 2\pi/3, \dots, 2\pi\}$. Therefore, by averaging over all rotation angles obtained with the random signal, we obtain $|f(\Delta_R,t)|$.

To reconstruct $|f(\Delta,t)|$, we vary the frequency $\Delta_R$ of the random signal within four separate frequency ranges $\Delta_R\in2\pi\times[0.118,0.558]$~MHz, $2\pi\times[0.642,1.104]$~MHz, $2\pi\times[1.2,1.752]$~MHz and $2\pi\times[1.850,2.352]$~MHz. For the CPMG scheme, the amplitude $b_R^\perp$ are $2\pi\times39$~kHz, $2\pi\times117$~kHz, $2\pi\times195$~kHz and $2\pi\times273$~kHz for the four ranges, respectively. For the GD$_\perp$ scheme, the corresponding amplitudes $b_R^\perp$ are $2\pi\times25$~kHz, $2\pi\times75$~kHz, $2\pi\times125$~kHz and $2\pi\times175$~kHz.

\section{The synchronized readout experiment} 

As shown in Fig. \ref{fig:1}, in the synchronized readout, a single frequency sensing procedure is repeatedly implemented at intervals of $T_L$ for the same continuous target signal. In the $m$-th sensing procedure of XY scheme, the state is firstly prepared into $\ket{+}$. Then, the corresponding control sequence is applied. The scan frequency is fixed at $2\pi \times 0.3$~MHz and the Rabi frequency of the control fields is $2\pi \times 10$~MHz. The initial phase of the signal slice (purple shaded region) in this single sensing procedure is
$\varphi_m = \omega_s m T_L + \varphi_0$. According to Eq.~(\ref{Eq:effective1_Appen}) in XY scheme, after sensing duration $T_s$, the accumulation phase of prepared state is 
\begin{equation}
\begin{split}
    \Phi_{\text{XY}}^{\text{syn}}
    &=\int_0^{T_s} b_s^\parallel F_{\text{XY}}(t^\prime)\cos(\omega_st^\prime+\varphi_m)\,dt^\prime\\
    &\approx\int_0^{T_s}\frac{2b_s^\parallel}{\pi}\cos(\delta t^\prime+\varphi_m)\,dt^\prime\\
    &=\frac{2b_s^\parallel}{\pi\delta}(\sin(\delta T_s+\varphi_m)-\sin(\varphi_m))\\
    & =\frac{2b_s^\parallel T_s}{\pi}\cos\left(\frac{\delta T_s}{2}+\varphi_m\right)\text{sinc}\left(\frac{\delta T_s}{2}\right),
\end{split}
\end{equation}
where $\delta=\omega_s-\omega_{\text{scan}}$, for $\delta T_s\ll1$, $\Phi_{\text{XY}}^{\text{syn}}\approx2b_s^\parallel T_s\cos(\varphi_m)/\pi$. Finally, a MW pulse implementing the operation $R_x(\pi/2)$ is applied prior to optical readout. During the optical readout ($\sim1 \mu$s), we detect an average of 0.09 photons and record the corresponding photon counts, denoted as $D_m$
\begin{equation}
    D_m\sim\text{Pois}[C(1-\epsilon \text{Bn}[P_L])],
\end{equation}
where $C$ is a variable readout gain and $\epsilon$ is the optical contrast. $\text{Pois}$ represents a Poisson process, and $\text{Bn}$ denotes a Bernoulli process, 
which takes the value 1 with probability $P_L$ and the value 0 with probability $1-P_L$. 
The probability of projecting the final state onto $\ket{L}$, denoted by $P_L$, is given by
\begin{equation}
    \begin{split}
P_{L}&=\frac{1}{2}\left[1+\sin(\Phi_{\text{XY}}^{\text{syn}})\right]\\
&=\frac{1}{2}\left[1+\sin\left(\frac{2b_sT_s}{\pi}\cos(\varphi_m)\right)\right].
    \end{split}
\end{equation}
We observe that the $m$-th measurement result depend on the initial phase $\varphi_m$. We achieve discrete-time sampling after M measurements. The frequency of target signal is obtained by performing a Fourier transform on the measurement results ensemble $\{D_m\}_{m=1}^M$.

In the $m$-th frequency sensing procedure of the GD$_\parallel$, CPMG and GD$_\perp$ scheme, the state is prepared into $\ket{+}$, $\ket{0}$ and $\ket{L}$, respectively. Then, the corresponding control sequence is applied. The scan frequency is fixed at $2\pi \times 0.3$~MHz and the Rabi frequency of the control fields is $2\pi \times 10$~MHz. Finally, a microwave pulse implementing the operation $R_x(\pi/2)$ and $R_y(3\pi/2)$ is applied for the GD$_\parallel$ and CPMG schemes, respectively, prior to optical readout. For the GD$_\perp$ scheme, a direct readout is performed without spin operation. 

According to Eqs.~(\ref{Eq:effective2_Appen}), (\ref{Eq:effective3_Appen}) and (\ref{Eq:effective4_Appen}), the probabilities of projecting final state onto $\ket{L}$ in GD$_\parallel$, onto $\ket{+}$ in CPMG, and onto $\ket{0}$ in GD$_\perp$ are given by 
\begin{subequations}
\begin{align}
&P_{L}=\frac{1}{2}\left[1+\sin\left(\frac{b_sT_L}{2}\cos(\varphi_m)\right)\right],\\ &P_{+}=\frac{1}{2}\left[1+\sin\left(\frac{2b_sT_L}{\pi}\cos(\varphi_m)\right)\right],\\
&P_{\ket{0}}=\frac{1}{2}\left[1+\sin\left(b_sT_L\cos(\varphi_m)\right)\right],
\end{align}
\end{subequations}
respectively.

In the synchronized readout experiment, the frequency components, amplitude and initial phase of continuous AC magnetic fields are consistent with magnetic fields used previously for frequency estimation. The pulse-sequence configuration is the same as that used for frequency estimation. For both the XY and GD$_\parallel$ schemes, the field is measured repeatedly with a sampling interval of $T_L = 71~\mu$s, and a 23-minute time trace of photon counts is recorded.
For both the CPMG and GD$_\perp$ schemes, the field is measured repeatedly with a sampling interval of $T_L = 31~\mu$s, and a 10-minute time trace of photon counts is recorded.

\begin{figure}[htbp]
\includegraphics[width=0.8\linewidth]{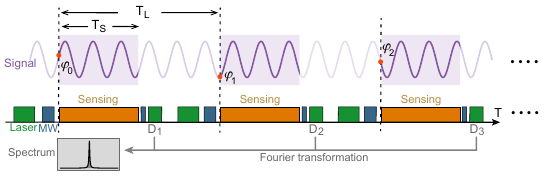}
\caption{The sequences of the synchronized readout technique.} \label{fig:1}
\end{figure}

\section{The experimental setup}
The NV center is hosted in a diamond substrate with less than 5~ppb N and natural isotope abundance, cut perpendicular to the [100] crystallographic direction. We use a home-built confocal microscope for the selective optical excitation and detection of fluorescence from single NV centers as shown in Fig.~\ref{fig:2}~(a). The excitation light~(532~nm) from a diode laser is digitally modulated by an acousto-optic modulator~(AOM) and then reflected to an oil-immersion objective lens~(NA=1.25) by a dichroic mirror. The photoluminescence (PL) from NV center is collected by the same microscope objective, which transmits the dichroic mirror and is detected by a single-photon detector. The static magnetic field ($\sim500$~Gs) is applied by a permanent magnet mounted on a three-dimensional translation stage. We determine the electron spin coherence time $T_2^{*}$ by performing a free induction decay (FID) experiment, obtaining $T_2^{*} = 9.3~\mu\mathrm{s}$. 

The MW control field and the high-frequency AC signals are generated by a microwave generator. The frequency, amplitude and phase of the MW fields are adjusted via in-phase and quadrature (IQ) modulation, and the arbitrary waveform generator (AWG) provides the corresponding I/Q waveforms. The low-frequency signals are directly generated from AWG. In the synchronized readout experiment, the signals in each sensing procedure are slices of a continuous signal, and the initial phases of these slices are time-correlated. By adjusting the sampling interval $T_L$, we ensure that the total duration of the ten sampling intervals is an integer multiple of the period of the target signal. The continuous target signals whose lengths exceed the AWG memory limit are obtained by repeating these ten measurement procedures. The MW field and AC signals are combined with a diplexer and delivered by a coplanar waveguide antenna. The antenna is glued to the diamond and connected to a printed circuit board (PCB). The PCB holder is attached on a $xyz$ piezo stage, which can be scanned by $70~\mu m\times 70~\mu m\times 70~\mu m$ to locate a single NV center. 
\begin{figure}[htbp]
\includegraphics[width=\linewidth]{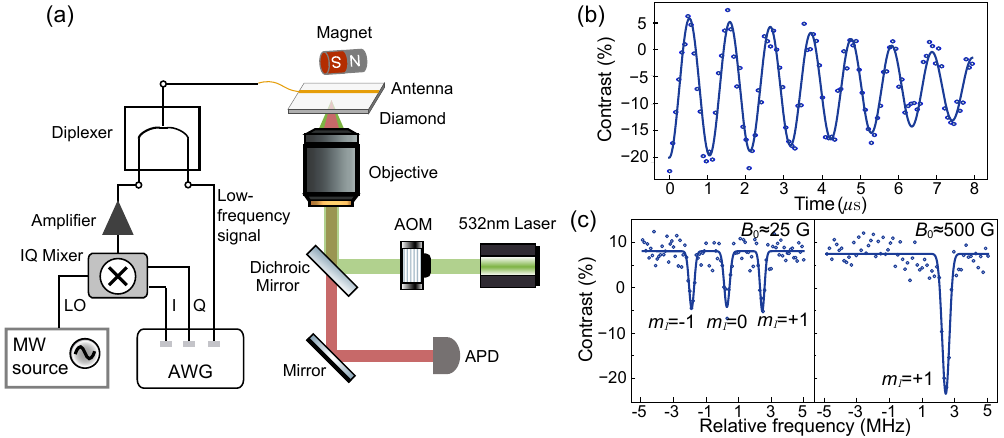}
\caption{(a) The illustration of home-built confocal microscope setup. (b) The result of the FID experiment for the electron spin is fitted with the function $0.5-0.5~\text{exp}(-t/T_2^*) \cos(vt)$. The coherence time is $ T_2^*=9.3\pm0.3\mu$s. (c) At a low static magnetic field ($\sim 25$~G), the nitrogen nuclear spin is unpolarized. At a higher static magnetic field of about $500$~G, the nuclear spin is polarized into the $m_I = +1$ state.} \label{fig:2}
\end{figure}
\end{widetext}

\bibliography{apssamp}
\end{document}